\documentclass[prx,twocolumn,floatfix,superscriptaddress,amsmath,aps,longbibliography]{revtex4-1}
\pdfoutput=1
\usepackage{dcolumn,graphicx,color,booktabs,microtype,afterpage}
\usepackage[colorlinks,plainpages=false,linkcolor=blue,urlcolor=blue,citecolor=blue,pdfpagemode=UseNone,pdfstartview=FitBH]{hyperref}

\usepackage{graphicx}
\usepackage{bmpsize}
\usepackage{amsfonts}
\usepackage{multirow}

\def\Rtm{$R\overline{3}m$}
\def\hoh{$(\frac{1}{2},0,\frac{1}{2})$}

\begin{document}
%\thispagestyle{empty}
%Notice: This manuscript has been authored by UT-Battelle, LLC under Contract No. DE-AC05-00OR22725 with the U.S. Department of Energy. The United States Government retains and the publisher, by accepting the article for publication, acknowledges that the United States Government retains a non-exclusive, paid-up, irrevocable, world-wide license to publish or reproduce the published form of this manuscript, or allow others to do so, for United States Government purposes. The Department of Energy will provide public access to these results of federally sponsored research in accordance with the DOE Public Access Plan (http://energy.gov/downloads/doe-public-access-plan).
%\pagebreak  \\
%\\
%\pagebreak \\
\title{Stripe antiferromagnetic ground state of ideal triangular lattice KErSe$_2$}
\author{Jie Xing}
\thanks{These authors contributed equally}
\affiliation{Materials Science and Technology Division, Oak Ridge National Laboratory, Oak Ridge, Tennessee 37831, USA}
\author{Keith M. Taddei}
\thanks{These authors contributed equally}
\affiliation{Neutron Scattering Division, Oak Ridge National Laboratory, Oak Ridge, TN 37831, USA}
\author{Liurukara D. Sanjeewa}
\thanks{These authors contributed equally}
\affiliation{Materials Science and Technology Division, Oak Ridge National Laboratory, Oak Ridge, Tennessee 37831, USA}
\author{Randy S Fishman} 
\affiliation{Materials Science and Technology Division, Oak Ridge National Laboratory, Oak Ridge, Tennessee 37831, USA}
\author{Marcus Daum}
\affiliation{School of Physics, Georgia Institute of Technology, Atlanta, Georgia 30332, USA.}
\author{Martin Mourigal}
\affiliation{School of Physics, Georgia Institute of Technology, Atlanta, Georgia 30332, USA.}
\author{C. dela Cruz}
\affiliation{Neutron Scattering Division, Oak Ridge National Laboratory, Oak Ridge, TN 37831, USA}
\author{Athena S. Sefat}
\affiliation{Materials Science and Technology Division, Oak Ridge National Laboratory, Oak Ridge, Tennessee 37831, USA}

\date{\today}

\begin{abstract}
Rare earth triangular lattice materials have been proposed as a good platform for the investigation of frustrated magnetic ground states.
KErSe$_2$ with the delafossite structure, contains perfect two-dimensional Er$^{3+}$ triangular layers separated by potassium ions, realizing this ideal configuration and inviting study. Here we investigate the magnetism of KErSe$_2$ at miliKelvin temperatures by heat capacity and neutron powder diffraction. Heat capacity results reveal a magnetic transition at 0.2 K in zero applied field. This long-range order is suppressed by an applied magnetic field of 0.5 T below 0.08 K.  Neutron powder diffraction suggests that the zero-field magnetic structure orders with $k=(\frac{1}{2},0,\frac{1}{2})$ in a stripe spin structure. Unexpectedly, Er is found to have a reduced moment of 3.06(1) $\mu_B$/Er in the ordered state and diffuse magnetic scattering, which originates at higher temperatures, is found to persist in the ordered state potentially indicating magnetic fluctuations. Neutron diffraction collected under an applied field shows a metamagnetic transition at $\sim$ 0.5 T to ferromagnetic order with $k$=(0,0,0) and two possible structures, which are likely dependent on the applied field direction. First principle calculations show that the zero field stripe spin structure can be explained by the first, second and third neighbor couplings in the antiferromagnetic triangular lattice. 

\end{abstract}

\maketitle

\section{Introduction}
The investigation of geometrically frustrated magnetism is intriguing due to the large ground state degeneracy it can engender~\cite{sadoc_mosseri_1999,Moessner}. Different arrangements of lattices and interaction types can give rise to a wide range of magnetic behavior, such as long-range antiferromagnetic ordering, spin-liquid physics,  spin-glass states or multi-critical phenomena where small perturbations in magnetic interactions or lattice symmetry have impact on the realized state~\cite{lacroix2011introduction}. In particular, quantum magnetic states - such as quantum-spin liquids (QSL) -  are especially interesting for their novel physics that may ostensibly open a new paradigm in computing. The QSL state generally requires an effective spin $S$=1/2 in frustrated (triangular) lattices. Such a configuration can manifest large spin fluctuations in the absence of long-range magnetic order (even down to zero Kelvin) and potentially lead to long-range entanged states whose excitations can exhibit non-abelian statistics ~\cite{Anderson,Mila2000,Balents}. 

Due to their possible Kramer doublets and spin-orbital coupling, rare earth compounds with odd 4$f$ electrons can be treated as $S_{\mathrm{eff}}$=1/2 at low temperatures and so are a potential avenue to realize QSL physics. For example, the triangular lattice compound YbMgGaO$_4$ has been proposed as having a QSL ground state~\cite{li2015gapless,li2015rare,li2016muon,shen2016evidence,paddison2017continuous,zhang2018hierarchy,li2017crystalline,steinhardt2019field}. However, more recently the intrinsic Mg/Ga disorder was found to perturb the system to other less-exotic magnetic states and so a suitable candidate material (preferably less prone to disorder) is still needed~\cite{li2015gapless,paddison2017continuous,zhang2018hierarchy,shen2018fractionalized,zhu2017disorder,zhu2018topography,kimchi2018valence}.

The delafossite structure \emph{A}\emph{R}\emph{Q}$_2$ (\emph{A}=Li, Na, K, Rb, Cs, Tl, Ag, Cu; \emph{R}=rare earth or 3/4$d$ transition metal; \emph{Q}= O, S, Se, Te) contains a perfect triangular \emph{R} site layered sublattice separated by spacer $A$ site ions. This structure and a receptiveness to magnetic ions makes it an ideal framework to search for quantum magnetism. One especially useful feature of the delafossite structure is that the intralayer and interlayer distances between magnetic ions can be tuned by using different spacer ions, potentially allowing for multiple magnetic ground states by changing the \textit{R}-site ion, or tuning three-dimensional and two-dimensional interactions. Thus far, promising results have been found in 3$d$ transition metal variants of the delafossite structure, which have had considerable interest, with some hints of exotic physics ~\cite{mitsuda1991neutron, kemp1990magnetic, borgers1966metamagnetism, komaba2010electrochemical, olariu2006unconventional, poienar2010spin, kadowaki1990neutron}. For example, CuFeO$_2$ can stabilize collinear commensurate, noncollinear incommensurate and collinear incommensurate magnetic structures alternatively via varying the strength of an applied magnetic field from 0 to 15 T \cite{mitsuda1991neutron,mitsuda1998partially}. 

When the magnetic ions are rare earths, the lattice is expanded, due to the large radius, and the interlayer/intralayer distances are increased causing further variability to the structure. For instance, Cs$R$Se$_2$ can crystallize in one of two different space groups $R\bar{3}m$ and \emph{P6$_3$/mmc} depending on the rare earth ion with both showing perfect rare earth triangular lattices~\cite{xing2019}. In the $A$YbSe$_2$ family, Yb$^{3+}$ hosts a $S_{\mathrm{eff}}$=1/2 and shows no long-range magnetic order down to the lowest measurement temperature, which may relate to the QSL ground state ~\cite{liu2018rare,baenitz2018naybs,ding2019gapless,ranjith2019field,bordelon2019field}. Furthermore magnetization measurements on single crystals of $A$YbSe$_2$ revealed several nearby long-range ordered states achievable by applying a magnetic field along different crystallographic directions ~\cite{xing2019field,ranjith2019anisotropic,ferreira2020frustrated,dai2020spinon,baenitz2018naybs}. On the other hand, in Ce based compounds, KCeS$_2$ and CsCeSe$_2$ exhibit the Ce$^{3+}$ 4$f^1$ electron configuration and a long-range order magnetic ground state was found in KCeSe$_2$ at $\sim$0.38 K~\cite{xing2019,KCeSe2}. In addition to these compounds, a spin glass state was proposed in the CsDySe$_2$ based on the AC magnetization measurement~\cite{xing2019}. 

Besides the Ce/Yb materials, Er based compounds are also attractive in the search for quantum magnetism and several compounds have already been discovered which seem promising. In ErMgGaO$_4$ there have been several reports of a possible QSL ground state with no apparent long-range order down to the lowest measured temperature~\cite{cai2020mu,cevallos2018structural}. Similarly, NaErSe$_2$, KErSe$_2$ and CsErSe$_2$ all exhibit easy-plane magnetic anisotropy with no long-range magnetic order or spin freezing in heat capacity and magnetization measurements down to 0.42 K~\cite{xing2019synthesis,xing2019,gao2019crystal}.
Interestingly, despite exhibiting no long-range order, both NaErSe$_2$ and KErSe$_2$ were found to exhibit a metamagnetic transitions under small magnetic fields when applied in the \emph{ab} plane ~\cite{xing2019synthesis}. Follow-up inelastic neutron scattering work studying the crystal electric field levels of KErSe$_2$, indicated the possibility of large anisotropic spin with low spin along $c$ axis~\cite{gao2019crystal,scheie2020crystal}. Such results together with the success of the similar Er triangular lattice material ErMgGaO$_4$, demand neutron diffraction measurements to study the magnetic properties down to mK temperatures in KErSe$_2$.

In this paper, we report heat capacity and neutron diffraction measurements performed at ultra-low temperatures and under applied magnetic fields on KErSe$_2$ together with first principles calculations to elucidate its magnetic behavior. Our heat capacity measurements show long-range magnetic order with an onset temperature of $\sim$0.2 K in zero field. Field dependent measurements show that the magnetic transition is suppressed with increasing field to below 0.08 K at 0.5 T. Additionally, at higher fields we observe a Schottky peak. Neutron powder diffraction measurements performed at 45 mK indicate a stripe-type magnetic structure at 0 T with ordering vector $k=(\frac{1}{2},0,\frac{1}{2})$ - a rare order for these systems. Under an applied magnetic field, the stripe-type order is found to be suppressed leading to a metamagnetic transition at $\sim$0.5 T to a ferromagnetic $k=(0,0,0)$ state. First principles calculations confirm that this magnetic state is possible in KErSe$_2$ but requires the inclusion of second and third nearest neighbor couplings, placing KErSe$_2$ in a unique region of the 112 exchange interaction phase diagram.  

\section{Materials and Methods}

A powder sample of KErSe$_2$ was synthesized by the solid state reaction, similar to reported previously in the literature\cite{xing2019synthesis}. Phase purity was checked with powder X-ray diffraction which found only the pure phase of KErSe$_2$ without any impurity peaks. The EDS analysis shows element molar ratio K:Er:Se is the expected stoichiometric ratio 1:1:2 within error. Temperature dependent heat capacity was measured in a Quantum Design Physical Properties Measurement System (PPMS) using the relaxation technique. 

Neutron powder diffraction measurements were performed on the HB-2A (POWDER) diffractometer of Oak Ridge National Laboratory's High Flux Isotope Reactor \cite{calder2018suite}. Diffraction patterns were collected using the open-21'-12' collimator settings (for pre-monochromator, pre-sample and pre-detector collimation respectively) and longer wavelength 2.41\AA\ incident wavelength setting. To achieve the desired temperature ranges and measure under an applied magnetic field an Oxford dilution refrigerator insert was used with a cryo-magnet allowing a temperature range of 0.03 - 300 K and applied fields between 0 and 5 T. Powder patterns were collected with count times of 4 hours per scan.  \lq Order-parameter\rq\ scans of peak intensities were collected by centering an individual detector on a peak position and collecting intensity on increasing temperature or applied field.

To ensure the sample achieved the desired mK base temperatures, a newly developed \lq over-pressurizable\rq\ sample can was used. To start, 3 g of KErSe$_2$ was pressed into a series of $\sim$ 0.5 g pellets and then rolled in Cu foil - thus preventing gain reorientation under the applied magnetic field. The foil-sample assembly was then loaded into a custom designed Cu powder can which included a small inlet in the lid to allow for pressurization after loading. The lid was sealed using In wire and the can was then connected to a pressure loader and pressurized with 10 atm of He gas. At 10 atm, the Cu tubing inlet (which was preloaded with In wire) was pressed closed in a hydraulic press and carefully cut. To ensure the 10 atm of He remained in the sample can, He leak checks were performed at each step of the loading process and again 24 hrs after the can was sealed.  Furthermore, as a check on the sample thermometry, powder patterns were collected continuously on cooling such that any nascent magnetic peaks could be monitored. Base temperature runs were not collected until the magnetic peak intensities remained constant for 4 hrs, ostensibly indicating base temperature was reached or that the moment had saturated. 

Analysis of the diffraction data was performed using the Rietveld method as implemented in the FullProf software suite \cite{Fullprof}. The Thompson-Cox-Hasting psuedo-Voight peak shape with axial divergence asymmetry was used to fit the instrumental profile of HB-2A \cite{Finger1994}. The magnetic structure analysis and solution was performed using the Simulated Annealing and Representational Analysis (SARAh) software \cite{Wills2000}. Visualization of the crystal structure was performed using VESTA \cite{Momma2011}.

\section{\label{sec:res} Results and Discussion}

\subsection{\label{subsec:HC} Heat Capacity}

Fig.~1 shows the temperature dependence of heat capacity of KErSe$_2$ from 1 K to 0.08 K. At high temperature, no long-range order was found in the previous report up to 9 T~\cite{xing2019synthesis}. At zero field, an apparent $\lambda$ shape anomaly was found at $\sim$0.2 K with a tail up to $\sim$0.4 K, revealing a second order phase transition in KErSe$_2$. The released magnetic entropy at the transition is close to 3 J/mol K. When the magnetic fields were applied along the $c$ axis, the long-range order was suppressed to low temperature and vanished at 0.5 T. This field-dependence is different from other rare earth triangular lattice materials~\cite{CeCd3P3,bordelon2019field}, possibly indicating a different magnetic structure. Additionally, a Schottky feature appeared above 0.5 T, indicating the potential mixture of the CEF states induced by the external magnetic field.

\begin{figure}[tb]
\includegraphics[width=1\linewidth]{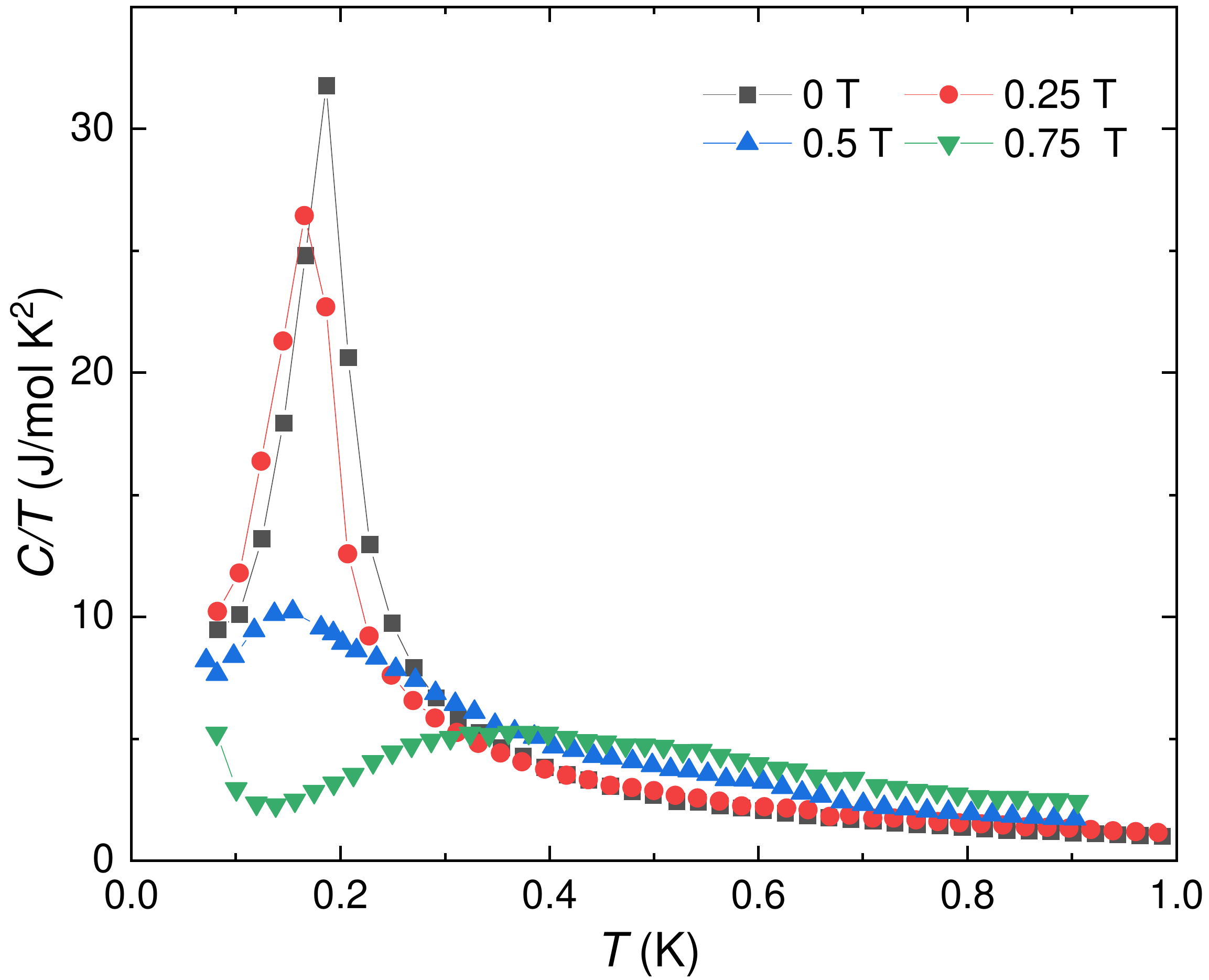}
\caption {Temperature dependence of heat capacity from 1 K to 0.08 K under different magnetic field.}
\label{ins_fig}
\end{figure}

\subsection{\label{subsec:neut}Neutron Scattering}

\begin{figure}
\includegraphics[width=\linewidth]{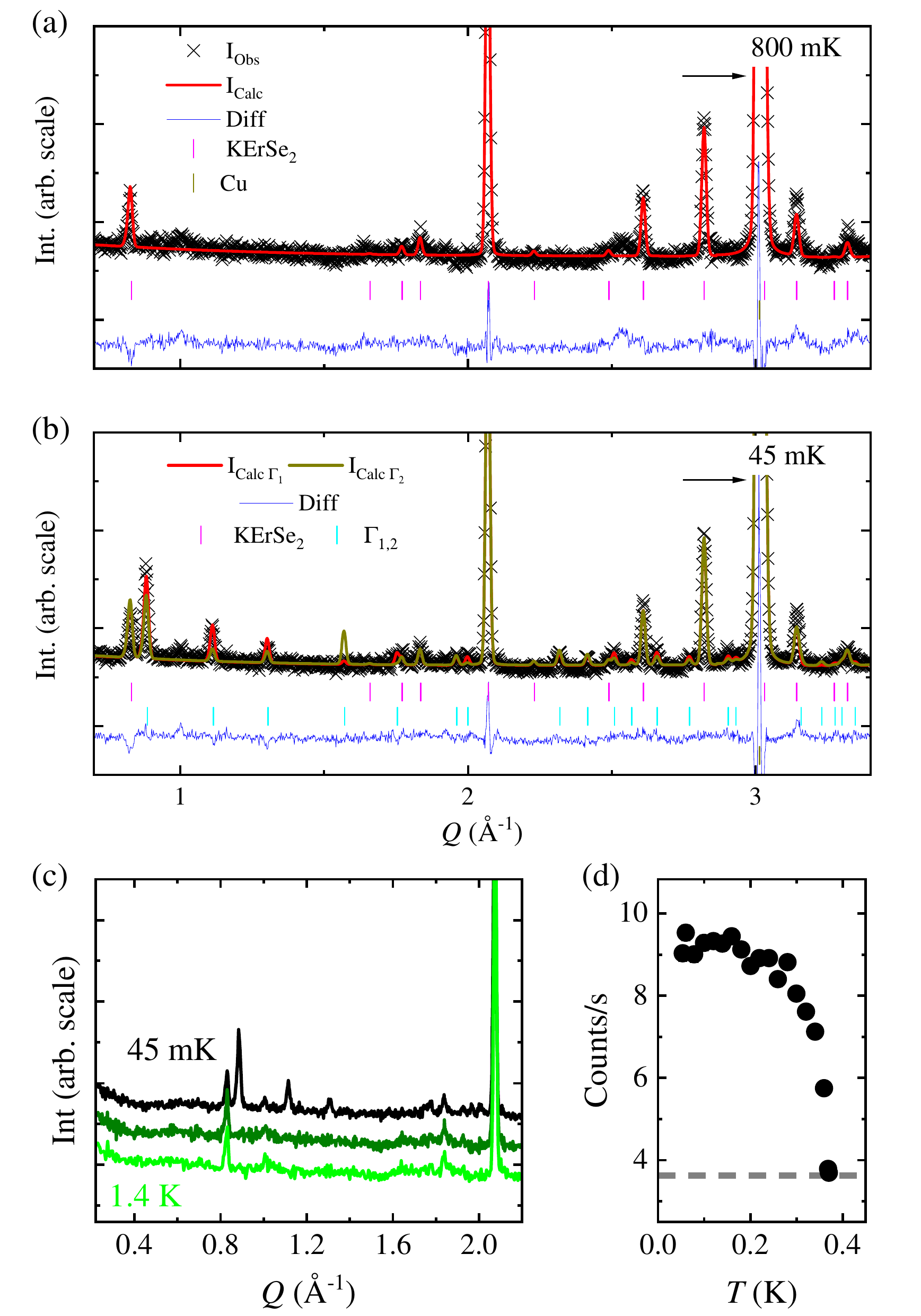}
\caption {\label{fig:fits} Neutron powder diffraction data, together with best fit models determined from Rietveld refinements for data collected at (a) 800 mK and (b) 45 mK. In both panels the data, model, difference, and phase peak indexes are indicated by black \lq $\times $\rq, red lines, blue lines, and magenta \lq $|$\rq  tick marks respectively. Indicated by dark yellow tick marks and asterisks are peaks belonging to the Cu sample can. In (b) the cyan tick marks indicate peaks belonging to the magnetic phase generated by the $\Gamma_{1}$ irreducible representation. (c) Neutron diffraction patterns collected at 1.4 K, 800 and 45 mK showing the appearance of new peaks below the transition temperature. (d) Temperature dependence of the $(\frac{1}{2},0,\frac{1}{2})$ peak intensity. In (b) the grey line indicates the background intensity for the $Q$ position of the $(\frac{1}{2},0,\frac{1}{2})$ peak determined from the 1.4 K diffraction pattern. }

\end{figure}

In order to investigate the nature of the long-range magnetic order, as indicated by the heat capacity measurements, neutron powder diffraction patterns were collected between 2 K and 0.08 K. A representative pattern collected above the signal seen in heat capacity (i.e. at 800 mK) is shown together with a best fit model determined by Rietveld refinement (Fig.~\ref{fig:fits}(a).) As seen, the previously reported delafossite crystal structure with \Rtm\ space group symmetry adequately accounts for the observed peak positions and intensities \cite{xing2019synthesis,scheie2020crystal}. Table~\ref{tab:crys} lists the crystallographic parameters for two selected temperatures obtained from Rietveld refinements.  

Figure~\ref{fig:fits} (c) shows neutron diffraction patterns collected at  1.4 K and 45 mK, which are above and below the feature temperature of the heat capacity result. As seen, at several positions new peaks arise in the 45 mK data indicating a symmetry breaking. Collecting the intensity of the peak at 0.89 \AA $^{-1}$ as a function of temperature (Fig.~\ref{fig:fits}(d)) reveals that the transition occurs at $\sim 370$ mK corresponding well with the transition temperature observed in the heat capacity measurements. Given the low scattering angle of these peaks, the low temperature of the transition and the form factor, we assume these peaks to be of a magnetic origin rather than due to a structural transition, and endeavor to model them using the tools of representational analysis for magnetic structure solution. 

\begin{table}
	\caption{\label{tab:crys}Crystallographic parameters of KErSe$_2$ at 370 and 45 mK. Parameters determined from Rietveld refinements performed using NPD collected on the HB-2A diffractometer with the 2.41 \AA\ incident wavelength. The atomic displacement parameters and magnetic moments are reported in units of \AA$^{-1}$ and $\mu_B/$Er, respectively. We note that the K and Er ions sit on the (0,0,0) $3a$ and (0,0,$\frac{1}{2}$) $3b$ Wyckoff sites, respectively, and so have no refinable component to their atomic positions. Only the Se has a refinable atomic position at the (0,0,$z$) $6c$ Wyckoff site. Atomic displacement parameters are not reported due to the optimization of the experiment for magnetic studies which precluded the high $Q$ data necessary for their reliable modeling.}
	\begin{ruledtabular}
		\begin{tabular}{llll}
     		 \multicolumn{1}{l}{\textit{T} } & \multicolumn{1}{c}{370 mK} & \multicolumn{1}{c}{45 mK}  \\
	\hline
	\multicolumn{1}{l}{Space Group} & \multicolumn{1}{c}{$R\overline{3}m$} & \multicolumn{1}{c}{$R\overline{3}m$}  \\
	\multicolumn{1}{l}{$R_{wp}$} & \multicolumn{1}{c}{ 6.38 \%} & \multicolumn{1}{c}{6.06 \%}  \\
	\multicolumn{1}{l}{$a$ (\AA)} & \multicolumn{1}{c}{4.1455(2)} & \multicolumn{1}{c}{4.1453(1)}  \\
	\multicolumn{1}{l}{$c$ (\AA)} & \multicolumn{1}{c}{22.711(1)} & \multicolumn{1}{c}{22.710(1)} \\
	\multicolumn{1}{l}{$V$(\AA$^3$)} & \multicolumn{1}{c}{338.01(3)} & \multicolumn{1}{c}{337.96(2)} \\
%	\multicolumn{1}{l}{K ($3a$)} 	&		&	\\
%	\multicolumn{1}{r}{$x$}	&	0	&	0	\\
%	\multicolumn{1}{r}{$y$}	&	0	&	0		\\
%	\multicolumn{1}{r}{$z$}	&	0	&	0		\\
	%\multicolumn{1}{r}{$U$} 	&	0.004(1)	&	0.002(2)	\\
	\multicolumn{1}{l}{Er ($3b$)} 	&		&		\\
%	\multicolumn{1}{r}{$x$}	&	0	&	0	\\
%	\multicolumn{1}{r}{$y$}	&	0	&	0	\\
%	\multicolumn{1}{r}{$z$}	&	0.5	&	0.5	\\
	%\multicolumn{1}{r}{$U$}	&	0.004(1)	&	0.002(2)		\\
	\multicolumn{1}{r}{$M$}	&	-	&	3.06(1)		\\
	\multicolumn{1}{l}{Se ($6c$)} 	&		&    		\\
%	\multicolumn{1}{r}{$x$}	&	0	&	0		\\
%	\multicolumn{1}{r}{$y$}	&	0	&	0		\\
	\multicolumn{1}{r}{$z$}	&	0.2351(2)	&	0.2354(2)		\\
	%\multicolumn{1}{r}{$U$}	&	0.004(1)	&	0.002(2)		\\	

		\end{tabular}
	\end{ruledtabular}
\end{table}

\begin{figure}
\includegraphics[width=\linewidth]{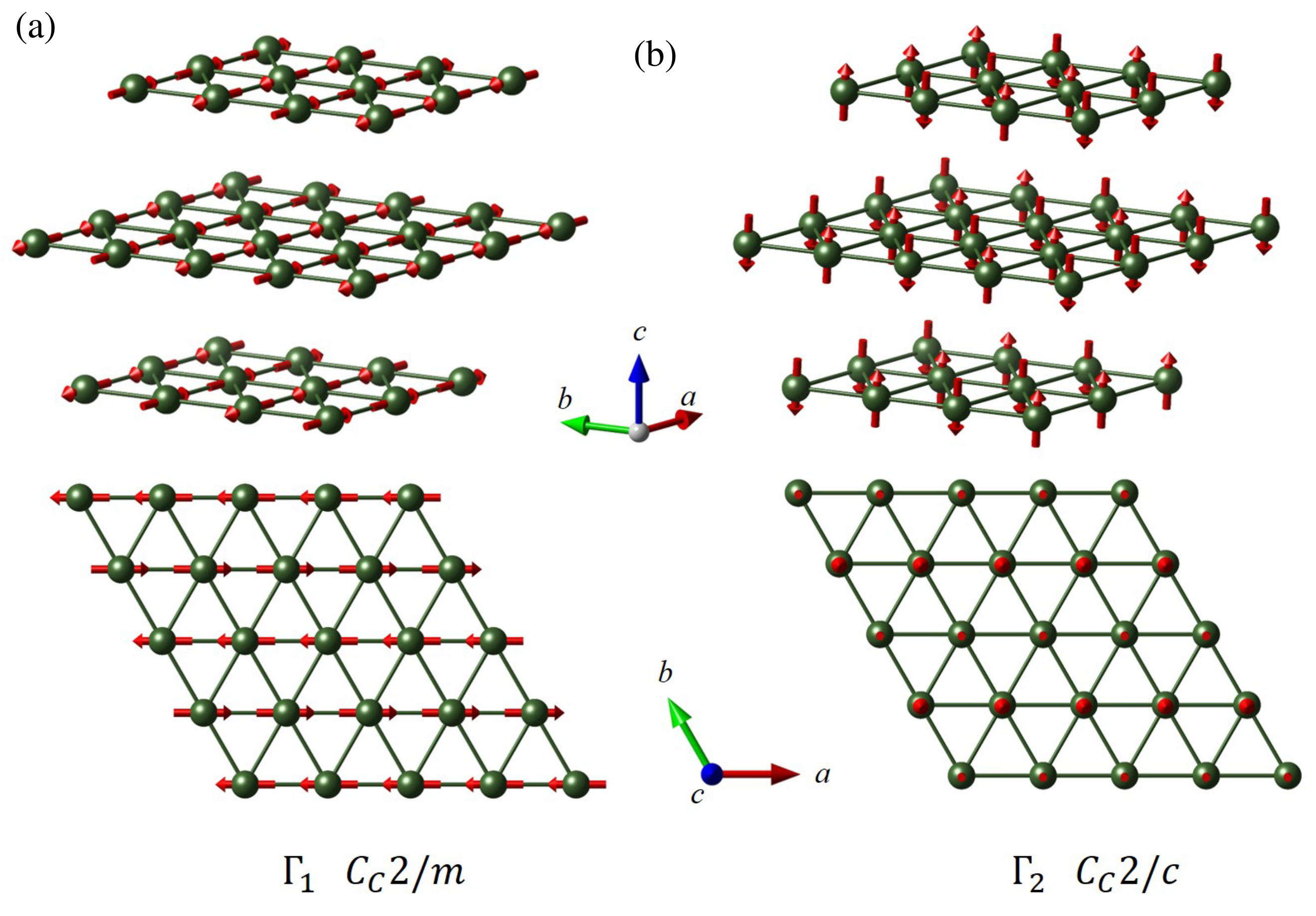}
\caption {\label{fig:magstruct1}Magnetic structures of the two irreducible representations formed from the decomposition of $R\overline{3}m$ with $k=(\frac{1}{2}, 0, \frac{1}{2})$. Along \textit{c} only half of the doubled magnetic unit cell is visualized in the upper two panels.}
\end{figure}

The appearance of new reflections at $Q$ positions distinct from the nuclear structure tells us that the magnetic order must have a non-zero ordering vector. We find that all of these additional peaks can be indexed by a \hoh\ $k$ vector and report in Table~\ref{tab:irreps} the decomposition of the \Rtm\ space group with the \hoh\ ordering vector for the $3b$ site. This analysis shows two irreducible representations ($\Gamma$) (corresponding to two magnetic space groups) are allowed, which together generate three possible basis vectors ($\Psi$). Considering the structures, we find that $\Gamma_{1}$ allows moments only in the \textit{a-b} plane along the \textit{b} axis, while $\Gamma_2$ has an in-plane component rotated away from either \textit{a} or \textit{b} and also (via $\Psi_3$) allows for a moment component along the \textit{c} axis. Representative magnetic structures for either irreducible representation are shown in Fig.~\ref{fig:magstruct1}.

%\begin{figure}
%\includegraphics[width=\linewidth]{zeroFieldMagOrder}
%\caption {\label{fig:magpeaks}(a) Neutron diffraction patterns collected at 1400, 800 and 45 mK showing the appearance of new peaks below the transition temperature. (b) Temperature dependence of the $(\frac{1}{2},0,\frac{1}{2})$ peak intensity. In (b) the grey line indicates the background intensity for the $Q$ position of the $(\frac{1}{2},0,\frac{1}{2})$ peak determined from the 1400 mK diffraction pattern.}
%\end{figure}

To discriminate between the two models, Rietveld refinements were performed using the zero-field data and the magnetic structure generated by each irreducible representation. The resulting best fit calculated diffraction patterns are shown together with the experimental data in Fig.~\ref{fig:fits}(b). As seen, $\Gamma_1$ visually reproduces the experimental data better, despite having fewer refinable parameters, and so we focus on that model for the zero-field structure. 

The fitted zero-field magnetic structure is shown in Fig.~\ref{fig:magstruct1}(a). As described the symmetry of $\Gamma_1$ only allows a non-zero magnetic moment along the \textit{b} axis. From our refinements, our extracted moment is 3.06(1) $\mu_B/$Er. We note that this is significantly less  than the $\sim 9.6 \mu_B/$Er expected for the Er$^{3+}$ ion, than the effective moment reported previously from magnetic susceptibility and other Er$^{3+}$ compounds with a similar valence~\cite{xing2019synthesis,taddei2019local}. There are several mechanisms which could cause this discrepancy: disorder leading to only partial ordering of the Er sites, strong frustration and(or) fluctuations preventing the full Er$^{3+}$ moment from ordering or issues with thermometry and thermal conduction leading to a non-saturated moment in our  base temperature run. For the latter, as described previously, we endeavoured to ensure thermalization at mK temperatures, and so we focus on the former two possibilities.  

%We now endeavor to work through these scenarios and discern which is most probable. To start, we consider sample thermalization. Recently, we developed a specially designed over-pressurized powder can to use with dilution refrigerators and powder samples~\cite{unpublished}. This technique validated the ability of powder samples to thermalize at mK temperatures in dilution refrigerators when loaded in copper cans charged with 10 atm of He. Furthermore, as a check on the sample thermometry, powder patterns were collected continuously on cooling such that the nascent magnetic peaks could be monitored. Base temperature runs were not collected until the magnetic peak intensities remained constant for 4 hrs, ostensibly indicating base temperature was reached or that the moment had saturated. As such we believe that the discrepancy in the moment size is inherent to the material's ground state magnetic order. 

\begin{table}
	\caption{\label{tab:irreps} Irreducible representations ($\Gamma$), constituent basis vectors ($\psi$), basis vector directions  and magnetic space groups for the $R\overline{3}m$ nuclear symmetry with $k = (\frac{1}{2}, 0, \frac{1}{2})$.}
	\begin{ruledtabular}
		\begin{tabular}{cccc}
     		 \multicolumn{1}{c}{$\Gamma$} & \multicolumn{1}{c}{$\psi$} & Components of $\psi$ & \multicolumn{1}{c}{Magnetic space group} \\
	\hline
	\multirow[t]{2}{*}{$\Gamma_1\ $} &  &  & $C_C2/m$     \\
									 & $\psi_1$  & $(0,-1,0)$  &          \\
	\multirow[t]{4}{*}{$\Gamma_2\ $} &   &  & $C_C2/c$     \\
									 & $\psi_2$  & $(2,1,0)$ &         \\
									 & $\psi_3$  & $(0,0,2)$ &         \\

		\end{tabular}
	\end{ruledtabular}
\end{table}

To discriminate among the fluctuation, frustration and disorder scenarios using our diffraction data we note that short-range static or fluctuating correlations can be observed in neutron powder diffraction patterns as background features. This occurs as short-range static correlations can cause broad diffuse features and fluctuations may be integrated into the diffraction pattern due to the finite energy of the incident neutrons and the diffractometer's inability to discriminate for purely elastic scattering. Looking at the background in Fig.~\ref{fig:fits}, we note that unlike the usually, flat isotropic background observed on the HB-2A diffractometer~\cite{calder2018suite}, we see a generally \lq lumpy\rq\ non-linear background in both the 800 and 45 mK scans. Such a signal may be indicative of diffuse scattering coming from the sample. 

\begin{figure}
\includegraphics[width=\linewidth]{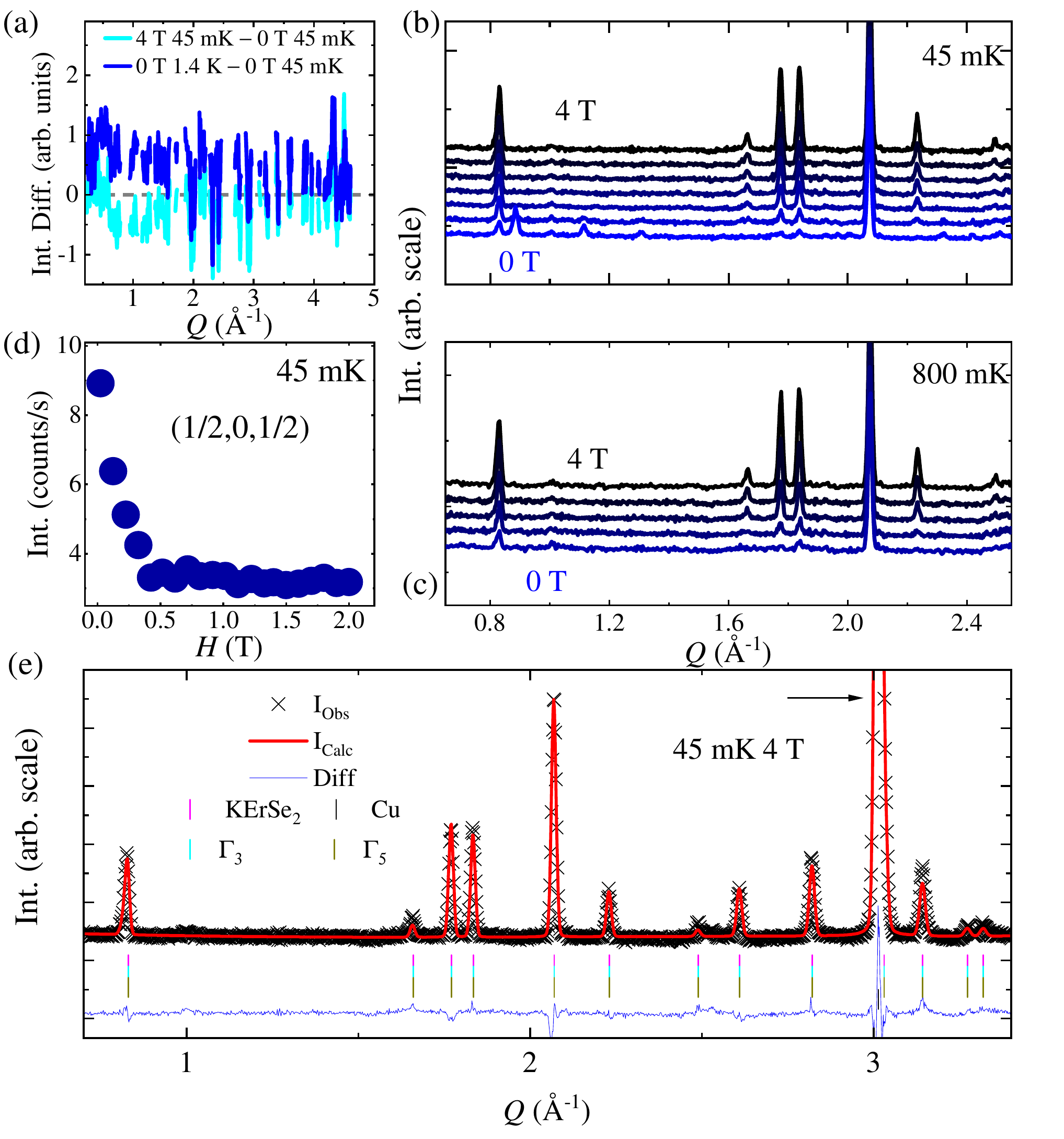}
\caption {\label{fig:Hpatterns} (a) Difference curves of diffraction patterns collected at various temperatures and fields. Diffraction patterns collected under applied fields at (b) 45 mK and (c) 800 mK. (d) Order parameter scan on the $(\frac{1}{2},0,\frac{1}{2})$ magnetic Bragg reflection as a function of applied magnetic field collected at 45 mK. (e) Neutron powder diffraction data, together with best fit models determined from Rietveld refinements for data collected at 45 mK under an applied field of 4 T. The data, model, difference, and phase peak indexes are indicated by black \lq $\times $\rq, red lines, blue lines, and \lq $|$\rq\  tick marks respectively. The Cu peaks are also indicated by an arrow due to the overlapping difference curve.}
\end{figure}

In order to look for sample dependence to this background we look at difference curves between the sample under different field and temperature conditions as shown in Fig.~\ref{fig:Hpatterns}(a).  Comparing first the 1.4 K and 45 mK data, we see a slight increase in the background at low $Q$ in the 1.4 K run as might be expected from the paramagnetic state where increased low $Q$ scattering is expected from the bare form factor of the Er$^{3+}$ ion (see for instance \onlinecite{taddei2019local} or \onlinecite{pajerowski2020quantification}). The lack of a significant feature here is consistent with the similarly \lq lumpy\rq\ background seen in both high and low temperature scans in Fig.~\ref{fig:fits} which indicates its source persists to 1.4 K. 

As another test, we compare low temperature patterns collected in 0 and 4 T applied fields (Fig.~\ref{fig:Hpatterns}(a)). In this difference curve a more significant change is apparent, and an obvious dip around 1 \AA$^{-1}$ is seen. As will be discussed later, both scans are in a magnetically ordered state and at the same temperature, therefore this cannot be due to paramagnetic scattering from Er$^{3+}$ ions. Instead, this is consistent with the applied field either polarizing static magnetically disordered regions into a FM state, or similarly, pushing fluctuating spins into long-lived (FM) states which contribute to Bragg reflections rather than diffuse scattering \cite{Sanjeewa2019dif}. Either of these explanations is consistent with the loss of diffuse intensity around 1 \AA$^{-1}$ upon applying a magnetic field. We note that similar difference curves are seen for all applied fields down to the lowest applied field of 0.25 T for which a full pattern was collected. 

From this observation we suggest it is likely that the reduced ordered moment for Er$^{3+}$ obtained from our Rietveld analysis is due to either partial static disorder where portions of the Er sub-lattice do not conform to the long range magnetic order, or to part of the Er$^{3+}$ ion's unpaired electrons remaining in a fluctuating state. While our current analysis is insufficient to discriminate between these scenarios decisively, we note that as only a small applied field is able to suppress the diffuse scattering the fluctuation scenario may be more likely as a larger field would be expected necessary to polarize static magnetic disorder. This is corroborated, at least in part, by recent studies explicating the magnetic Hamiltonian which suggested the possibility of quantum fluctuations in the magnetic ground state~\cite{scheie2020crystal}. However, additional studies on single crystals are needed to look for firmer evidence of diffuse scattering and its potential features in un-collapsed \textit{HKL} space as well as measuring the spin-waves in the ordered state.

%\begin{figure}
%\includegraphics[width=\linewidth]{FieldFit}
%\caption {\label{fig:fieldfit} Neutron powder diffraction data, together with best fit models determined from Rietveld refinements for data collected at 45 mK under an applied field of 4 T. The data, model, difference, and phase peak indexes are indicated by black $\lq \times \rq$, red lines, blue lines, and $\lq | \rq $ tick marks respectively.}
%\end{figure}

Encouraged by our heat capacity results, we performed neutron powder diffraction measurements under applied magnetic fields to look for a change in the long-range magnetic order. Powder patterns collected at 45 mK and 800 mK under applied fields from 0 to 4 T are shown as waterfall plots in Fig.~\ref{fig:Hpatterns} (b) and (c). At 45 mK, we observe the relatively quick suppression of the \hoh\ type magnetic state, with the associated magnetic Bragg peaks disappearing above 0.25 T. On the other hand, while peaks associated with the \hoh\ order are suppressed, a significant increase in intensity is seen on several low $Q$ nuclear Bragg peaks, which appears to saturate above 0.5 T indicating the stabilization of a new magnetic order with $k=(0,0,0)$. Performing an order parameter scan as a function of applied field on the $(\frac{1}{2},0,\frac{1}{2}$) peak (Fig.~\ref{fig:Hpatterns}(d)), we see its intensity is fully suppressed by 0.5 T. This together with the change in magnetic scattering to nuclear peak positions indicates a metamagnetic transition. This finding is consistent with the previous reported isothermal magnetization results~\cite{xing2019synthesis}.

Using $k=(0,0,0)$ we again perform representational analysis to identify the possible magnetic structures under applied field, the results of which are shown in Table~\ref{tab:irreps2}. Again only two $\Gamma$ are found with a total of three $\Psi$. In this case the first irreducible representation ($\Gamma_{3}$) allows only a moment along the \textit{c} axis, while the second ($\Gamma_5$) has in-plane moments with one $\Psi$ along \textit{b} and the other rotated between the two in-plane lattice vectors. 

Looking at the 800 mK waterfall plot (Fig.~\ref{fig:Hpatterns}(c)), we find that even above the zero-field $T_N$ the applied field is able to induce long-range magnetic order - in accord with the heat capacity measurements. As seen the $k=0$ type peaks, at positions where the nuclear intensities are otherwise weak, are clearly visible by 1 T. These peaks are identical in position and relative intensity to those of the low temperature $k=0$ magnetic structure described above and so we suggest are an extension of that magnetic order to higher temperatures.

\begin{table}
	\caption{\label{tab:irreps2} Irreducible representations ($\Gamma$), constituent basis vectors ($\psi$), basis vector directions  and magnetic space groups for the $R\overline{3}m$ nuclear symmetry with $k = (0,0,0)$.}
	\begin{ruledtabular}
		\begin{tabular}{cccc}
     		 \multicolumn{1}{c}{$\Gamma$} & \multicolumn{1}{c}{$\psi$} & Components of $\psi$ & \multicolumn{1}{c}{Magnetic space group} \\
	\hline
	\multirow[t]{2}{*}{$\Gamma_3\ $} &  &  & $R\overline{3}m'$     \\
									 & $\psi_1$  & $(0,0,12)$  &          \\
	\multirow[t]{4}{*}{$\Gamma_5\ $} &   &  &      \\
									 & $\psi_2$  & $(0,-3,0)$ &  $C2/m$        \\
									 & $\psi_3$  & $(-3.5,-1.7,0)$ & $C2'/m'$         \\

		\end{tabular}
	\end{ruledtabular}
\end{table}

While it is generally suspect to attempt magnetic structure solution for powder data in applied fields due to the lack of control over the field orientation with respect to the crystal lattice, considering the few possible structures consistent with $k=0$ we attempt such an analysis nonetheless. Doing so we find that neither allowed $\Gamma$ alone is able to produce an adequate fit to the data. However, if both $\Gamma$ are used as two separate magnetic phases we find we are able to reproduce the observed intensities quite well using only one $\Psi$ from each $\Gamma$ ($\Psi_1$ and $\Psi_3$) (Fig.~\ref{fig:Hpatterns} (e).) This describes a state where the applied field pushes different grains of the powder sample into different magnetic structures ostensibly depending on their relative orientation with respect to the applied field (e.g.grains with the field mostly in-plane may lead to one structure while those with the field mostly out-of-plane may lead to another.) As the \hoh\ type peaks are completely suppressed we can surmise that the sample must be in one of three states: one of the two $k=0$ type magnetic structures or a PM state. As refinements of the magnetic phases rely on intensities arising purely from the magnitude of the magnetic moment, it is not possible to uniquely determine phase fractions or magnetic moments for either structure nor whether the entire sample is ordered or whether some of it has become PM. Such questions will require follow up diffraction studies performed with applied fields on a single crystal sample. 

\begin{figure}
\includegraphics[width=\linewidth]{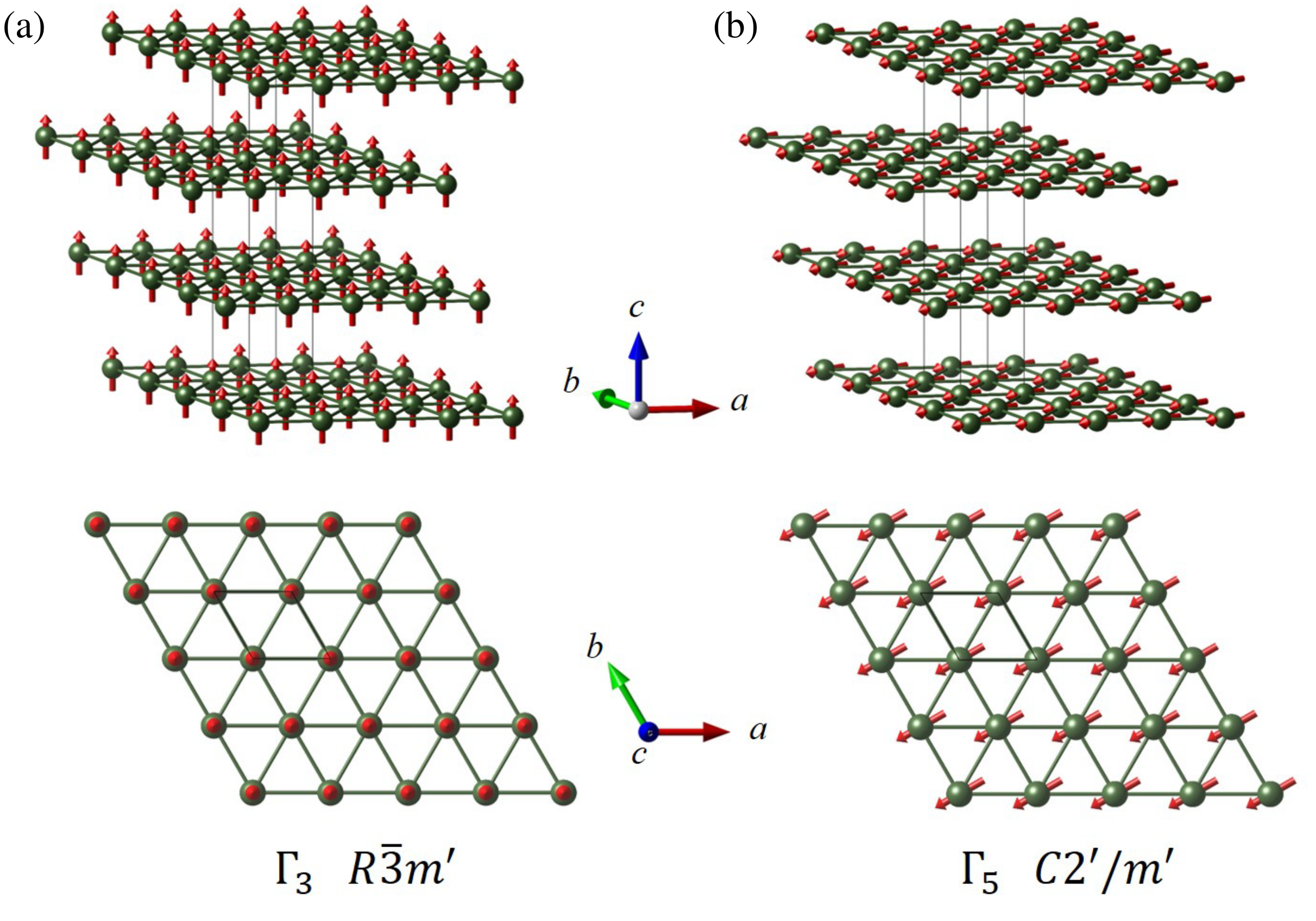}
\caption {\label{fig:magstruct}Magnetic structures of the two irreducible representations formed from the decomposition of $R\overline{3}m$ with $k=(0, 0, 0)$.}
\end{figure}

However, from our work we can say that an applied magnetic field drives a metamagnetic transition from the \hoh\ type order to a FM $k=0$ state. Additionally, our analysis shows that one of two structures is adopted (likely dependent on the field's orientation with respect to the crystal lattice) with one being a fully \textit{c} polarized FM state (Fig.~\ref{fig:magstruct}(a)) and the other a FM structure with the moments in-plane (Fig.~\ref{fig:magstruct}(b)). While the former may be more surprising due to the expected easy-plane type single ion physics in this material, the latter is less so, and ostensitly indicates the weak inter-layer correlations which may be suppressed with a 0.5 T field from the AFM inter-layer correlations found in the 0 T structure to the currently discussed FM in-plane structure.

\subsection{\label{subsec:TM} Theory Model}

\begin{figure}
\includegraphics[width=8.5cm]{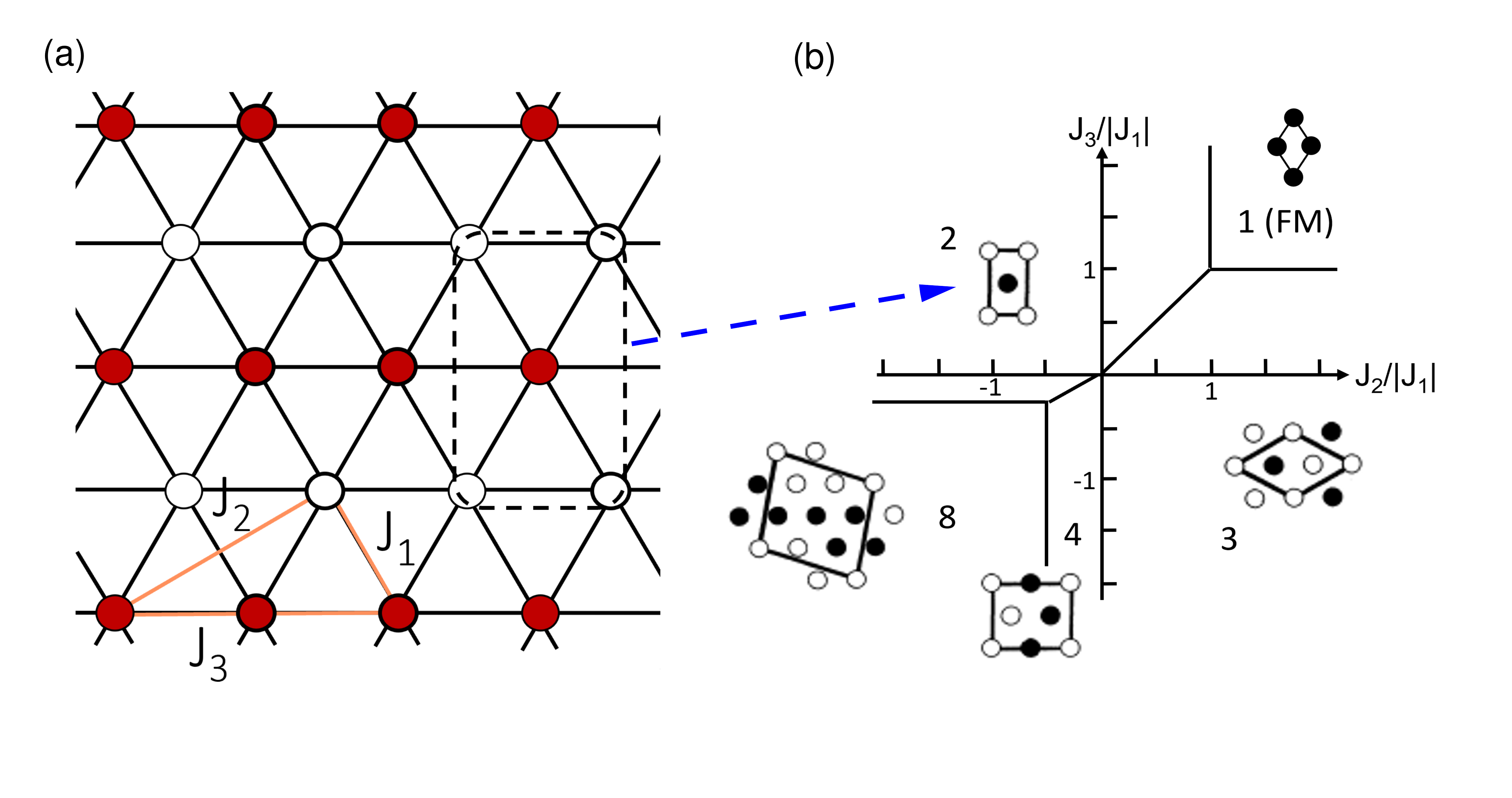}
\caption{\label{fig:theory} (a) Triangular lattice with first, second, and third neighbor couplings.  The 2-SL magnetic unit cell is given by the dashed boundary. (b) Phase diagram of the triangular lattice showing the stable magnetic phases with interactions $J_1$, $J_2$, and $J_3$.  Only collinear spin states are 
considered.}
\end{figure}

Based on the structure, we proposed a simple two-dimensional Hamiltonian for KErSe$_2$: 
\begin{equation}
{\cal H} = -\sum_{i,j} J_{ij} S_i \cdot S_j - K \sum_i {S_{ix}}^2,
\end{equation}
where sites $i$ and $j$ on a triangular lattice are coupled by exchange $J_{ij}$.  When the nearest-neighbor coupling $J_1$ is negative, this system is magnetically frustrated.
However, this frustration can be lifted by second and third-neighbor exchange couplings $J_2$ and $J_3$.  All three exchange couplings are shown in Fig.~\ref{fig:theory}(a).

The spin state observed in KErSe$_2$ has wavector ${\bf Q}=(2\pi/a)(1,0)$ in the $xy$ plane.  In terms of the reciprocal lattice vectors
\begin{equation}
{\bf b}_1=\frac{2\pi }{a} (1,1/\sqrt{3}),
\end{equation}
\begin{equation}
{\bf b}_2=\frac{2\pi }{a} (1,-1/\sqrt{3}),
\end{equation}
we can write ${\bf Q}=(1/2,1/2)\equiv ({\bf b}_1+{\bf b}_2)/2$.  The spin state is sketched in Fig.~\ref{fig:theory}(b), where dark circles mean the spin points to the right and open circles mean the spin points to the left.
This state has two spins in the magnetic unit cell, which is shown as the dashed boundary.
In the previous study in the antiferromagnetic triangular lattice~\cite{J3}, the observed two-sublattice (2-SL) state could be stable over a wide range of $J_2/\vert J_1\vert $ and
$J_3/\vert J_1\vert $, as shown in Fig.~\ref{fig:theory}(b). KErSe$_2$ is a possible representative of this stripe spin state in the delafossite structure.

Previously, long-range magnetic order was reported in two rare earth delafossite materials. The Ce based compounds present a long-range order at zero field, while Yb based materials present spin liquid ground state and need an applied magnetic field to induce the long-range order~\cite{KCeSe2,xing2019field,ranjith2019field,bordelon2019field}. Although the magnetic transition temperature is close in Ce and Er based materials, the long-range orders show different field dependent behavior. The dome feature was found in $H-T$ phase diagram of Ce compounds~\cite{KCeSe2}. In KErSe$_2$ the magnetic transition monotonously decreases to lower temperature with increasing the magnetic field. The significant difference between these results indicates the possible multiple magnetic ground states in the similar magnetic triangular lattice with different rare earth ions. 

On the other hand, although the specific rare earth ions play an important role in the similar delafossite structure, we could also tune the intralayer and interlayer distance by replacing the nonmagnetic ions and may find the boundaries of different spin states. Based on the experiments of Er and Yb based samples, the replacement of the nonmagnetic ions in rare earth delafossites show a slight influence on the magnetic properties~\cite{xing2019,ranjith2019field,xing2019synthesis}. In the predicted phase diagram from theory,  multiple ground states could be explored by changing the strength of interaction\cite{maksimov2019anisotropic}. Our KErSe$_2$ result could help to understand the phase diagram and discover the spin liquid state in the rare earth compounds.

\section{Conclusion}

In conclusion, we used heat capacity and neutron powder diffraction to investigate the magnetism of KErSe$_2$  in the low temperature region. The heat capacity reveals a long-range magnetic order at $\sim$0.2 K at zero field with a long tail up to $\sim$0.4 K. A magnetic field is found to suppress the magnetic transition to below 0.08 K at 0.5 T. Neutron powder diffraction results suggest the zero field magnetic structure is an unusual stripe spin structure with a reduced moment of 3.06(1) $\mu_B$/Er. Analysis of the powder patterns' \lq background\rq\ features reveals the presence of a magnetic diffuse signal which may indicate fluctuating spin even in the ordered state consistent with the observed reduced ordered moment. Diffraction data collected under an applied field confirm the heat capacity results and indicate a metamagnetic transition from AFM to FM. Additionally, careful analysis shows that the metamagnetic transition is likely dependent on the direction of the applied field with respect to the crystal lattice leading to at least two possible field stabilized FM orders and a likely complex temperature/field magnetic phase diagram with several phases at similar energy scales. Using first principles calculations we find that two-dimensional Hamiltonian which considers the first, second and third neighbor couplings can explain this magnetic structure.
Our results could help to understand the phase diagram of the rare earth triangular lattice materials.
%\newline
%\newline

\section*{Acknowledgments}

We would like to thank M. Cochran for his work implementing and preparing the high He pressure powder cans which enabled our powder sample to reach the dilution refrigeration temperatures in the time scale of a neutron scattering experiment. Additionally, we thank M. Rucker for his help overpressurizing the powder can. 
The research at the Oak Ridge National Laboratory (ORNL) is supported by the U.S. Department of Energy (DOE), Office of Science, Basic Energy Sciences (BES), Materials Sciences and Engineering Division.
The part of this research conducted at ORNL's High Flux Isotope Reactor was sponsored by the Scientific User Facilities Division, Office of BES, U.S. DOE.
The work at Georgia Tech (Heat Capacity Measurements) was supported by the National Science Foundation through Grant No. NSF- DMR-1750186.

This manuscript has been authored by UT-Battelle, LLC under Contract No. DE-AC05-00OR22725 with the US Department of Energy. The United States Government retains and the publisher, by accepting the article for publication, acknowledges that the United States Government retains a nonexclusive, paid-up, irrevocable, worldwide license to publish or reproduce the published form of this manuscript, or allow others to do so, for United States Government purposes. The Department of Energy will provide public access to these
results of federally sponsored research in accordance with the DOE Public Access Plan(http://energy.gov/downloads/doe-public-access-plan).
%
%\end{document}

\end{document}